\documentclass{webofc}

\usepackage[varg]{txfonts}   
\usepackage{hyperref}
\usepackage{url}
\usepackage{amsmath}
\usepackage{subfigure}
\usepackage{physics}

\hypersetup{colorlinks=true,citecolor=blue,urlcolor=blue,linkcolor=blue}
\begin{document}
\title{$\mu^-\mu^+\to {\nu_\mu}\bar{\nu}_\mu t\bar{t}H$ amplitudes in the Feynman-diagram gauge}

\author{\firstname{Ya-Juan} \lastname{Zheng}\inst{1}\fnsep\thanks{\email{yjzheng@iwate-u.ac.jp}} 
}

\institute{Faculty of Education, Iwate University, Morioka, Iwate 020-8550, Japan }

\abstract{We study the process $\mu^-\mu^+\to {\nu_\mu}\bar{\nu}_\mu t\bar{t}H$ with complex CP violating $ttH$ couplings in the SMEFT with a dimension-6 operator. 
When the amplitudes are expressed in the Feynman-Diagram gauge, the dominance of the total cross section via the weak boson fusion diagrams is manifest. The high energy behaviour is dictated by the higher-dimensional vertices in the dimension-6 SMEFT operator. These properties are not manifest in the unitary gauge because of subtle cancellation among diagrams. 
}
\maketitle
%
\label{intro}

Feynman-diagram (FD) gauge propagator is proposed first for massless gauge bosons~\cite{Hagiwara:2020tbx}, and then for massive gauge bosons~\cite{Chen:2022gxv,Chen:2022xlg},
where the gauge vector of the light-cone gauge is chosen along the opposite of the gauge boson three-momentum direction:
\begin{eqnarray}
    n(q)_{\rm FD}^\mu
    =\left(
    {\rm sgn}(q^0),-\vec{q}/|\vec{q}|
    \right)
    \label{eq:nqFD}.
\end{eqnarray}

In QED and QCD, scattering amplitudes in the FD gauge are obtained simply by replacing the photon and the gluon propagators as 
\begin{eqnarray}
    G_{\mu\nu}(q)
    =\frac{i}{q^2+i\epsilon}
    \left(
    -g_{\mu\nu}+\frac{n_\mu q_\nu+q_\mu n_\nu}{n\cdot q}
    \right),
\end{eqnarray}
in the numerical codes of HELAS~\cite{Hagiwara:1990dw,Murayama:1992gi}.
In this work~\cite{Hagiwara:2024xdh}, we extend this prescription into the electroweak (EW) sector by introducing new Feynman rules which take the Goldstone bosons as the fifth component of the massive gauge bosons for arbitrary gauge model. As the key step, the massive gauge bosons wave function in the FD gauge $\epsilon^M (p,\lambda)$, ($M=\mu,4$) requires one more component than the usual four component wave function $\epsilon^\mu(p,\lambda)$ since the Goldstone boson component is included as a part of the physical weak boson in the FD gauge. The explicit form for the transverse and longitudinal components are
\begin{eqnarray}
\epsilon^M(p,\pm1)=\left(\epsilon^\mu(p,\pm1),\,0\right),
\quad\quad
\epsilon^M(p,0)=\left(\tilde{\epsilon}^\mu(p,0),\,i\right),
\end{eqnarray}
where the reduced polarization vector is defined as 
\begin{eqnarray}
\tilde{\epsilon}^\mu(p,0)=\epsilon^\mu(p,0)-\frac{p^\mu}{m}
=
-\frac{m\, n(p)^\mu_{\rm FD}}{p\cdot n(p)_{\rm FD}}
.
\end{eqnarray}
In terms of the light-cone gauge vector~\eqref{eq:nqFD}, the $5\times5$ weak boson propagator can be expressed as~\cite{Chen:2022xlg}
\begin{eqnarray}
    G_{MN}(q)=
    \frac{i}{q^2-m^2+i\epsilon}
\begin{pmatrix}
  -g_{\mu\nu}+\frac{q_\mu n_\nu+n_\mu q_\nu}{n\cdot q} & i\frac{mn_\mu}{n\cdot q}\\
  -i\frac{mn_\nu}{n\cdot q} & 1
\end{pmatrix}.
\label{eq:GMN}
\end{eqnarray}
The above FD gauge propagator contains not only the four-vector and the Goldstone propagation terms, $(\mu,\nu)$ and $(4,4)$ components, respectively, but also the mixing between the 4-vector and the Goldstone components. In ref.\,\cite{Hagiwara:2024xdh}, we introduce new Feynman rules to generate automatically the FD gauge amplitudes, and successfully implemented into the Madgraph~\cite{Alwall:2014hca}, which will be released in version 3.6.0.

As a demonstration of the above implementation, we study the muon collider process $\mu^-\mu^+\to \nu_\mu\bar{\nu}_{\mu}\bar{t}tH$ in the framework of SMEFT with a dimension-6 operator which modifies the top Yukawa coupling as:
\begin{eqnarray}
    {\cal L}_{\rm SMEFT}={\cal L}_{\rm SM}+
    \left\{\frac{\lambda}{\Lambda^2}\left(Q_3^\dagger\tilde{\phi} t_R\right)\left(\tilde{\phi}^\dagger\tilde{\phi}-\frac{v^2}{2}\right)+{\rm h.c.}\right\},
    \label{eq:LSMEFT}
\end{eqnarray}
where $Q_3=\left(t_L,b_L\right)^T$ and 
\begin{eqnarray}
    \tilde{\phi}=\left(\frac{v+H-i\pi^0}{\sqrt{2}},-i\pi^-\right).
\end{eqnarray}
The above SMEFT Lagrangian gives the CP-violating top-Yukawa coupling,
\begin{eqnarray}
    {\cal L}_{ttH}=-g\bar{t}(\cos\xi+i\sin\xi\gamma_5)tH,
\end{eqnarray}
when the coefficient of the dimension-6 operator is taken as 
~\cite{Barger:2023wbg} 
\begin{eqnarray}
    \frac{\lambda}{\Lambda^2}
    =\frac{\sqrt{2}(g_{\rm SM}-ge^{i\xi})}{v^2}.
\end{eqnarray}
In the numerical calculation, we set $g=g_{\rm SM}=\frac{m_t}{v}$.

For the ${\nu_\mu}\bar{\nu}_\mu ttH$ production, when we generate the process through Madgraph, the Feynman diagrams can be categorized into the following groups: 
\begin{itemize}
    \item WWF: $W^-W^+$ fusion,
    \item $W\mu+\mu W$: $W^-\mu^+$ fusion and $\mu^-W^+$ fusion, 
    \item anni.: $\mu^-\mu^+$ annihilation with $s$-channel $Z,\gamma$ or $t$-channel $\mu$ exchange.
\end{itemize}

\begin{figure}[t]
\centering
\subfigure[]{
\includegraphics[width=5cm,clip]{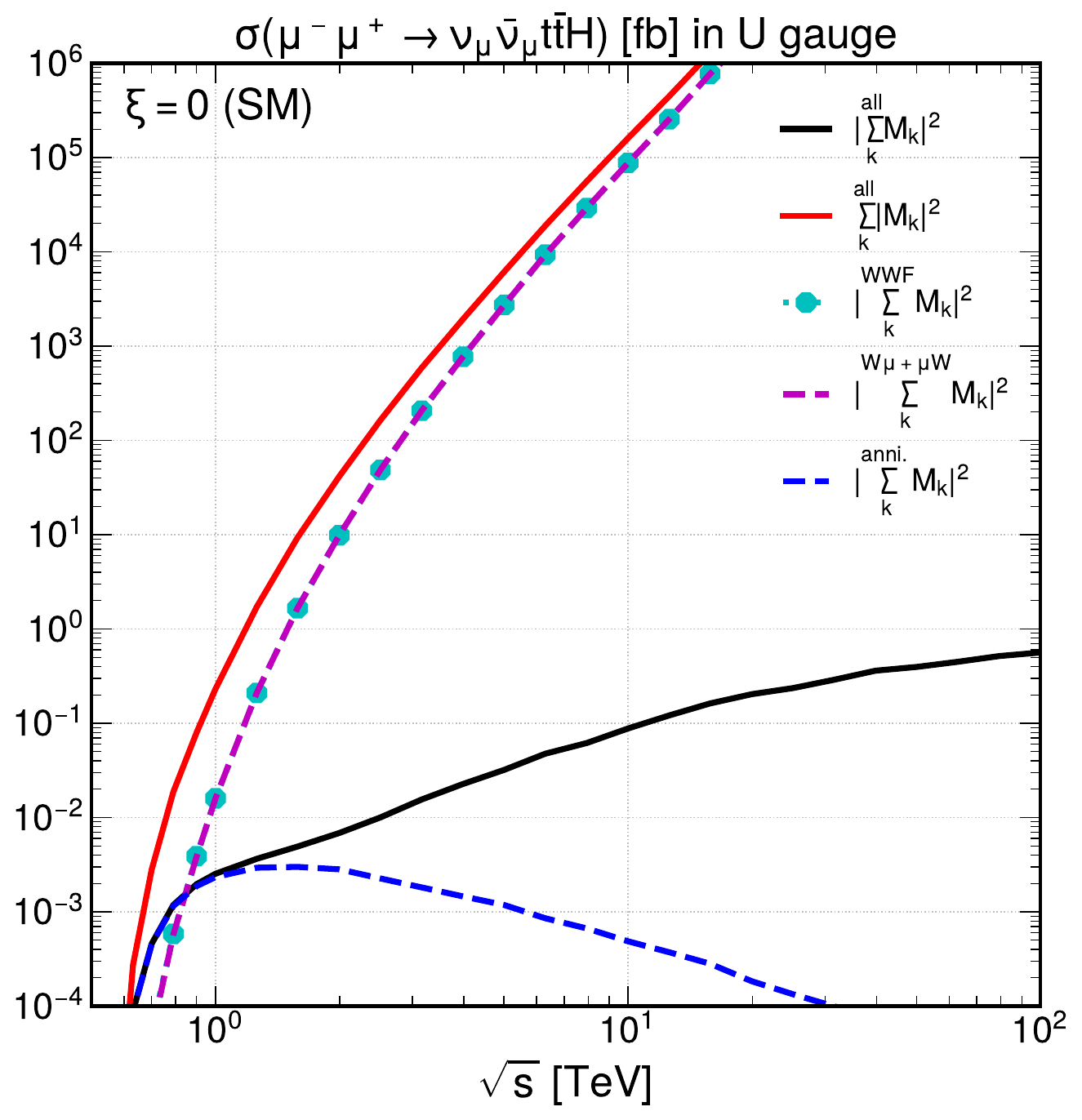}}
\subfigure[]{
\includegraphics[width=5cm,clip]{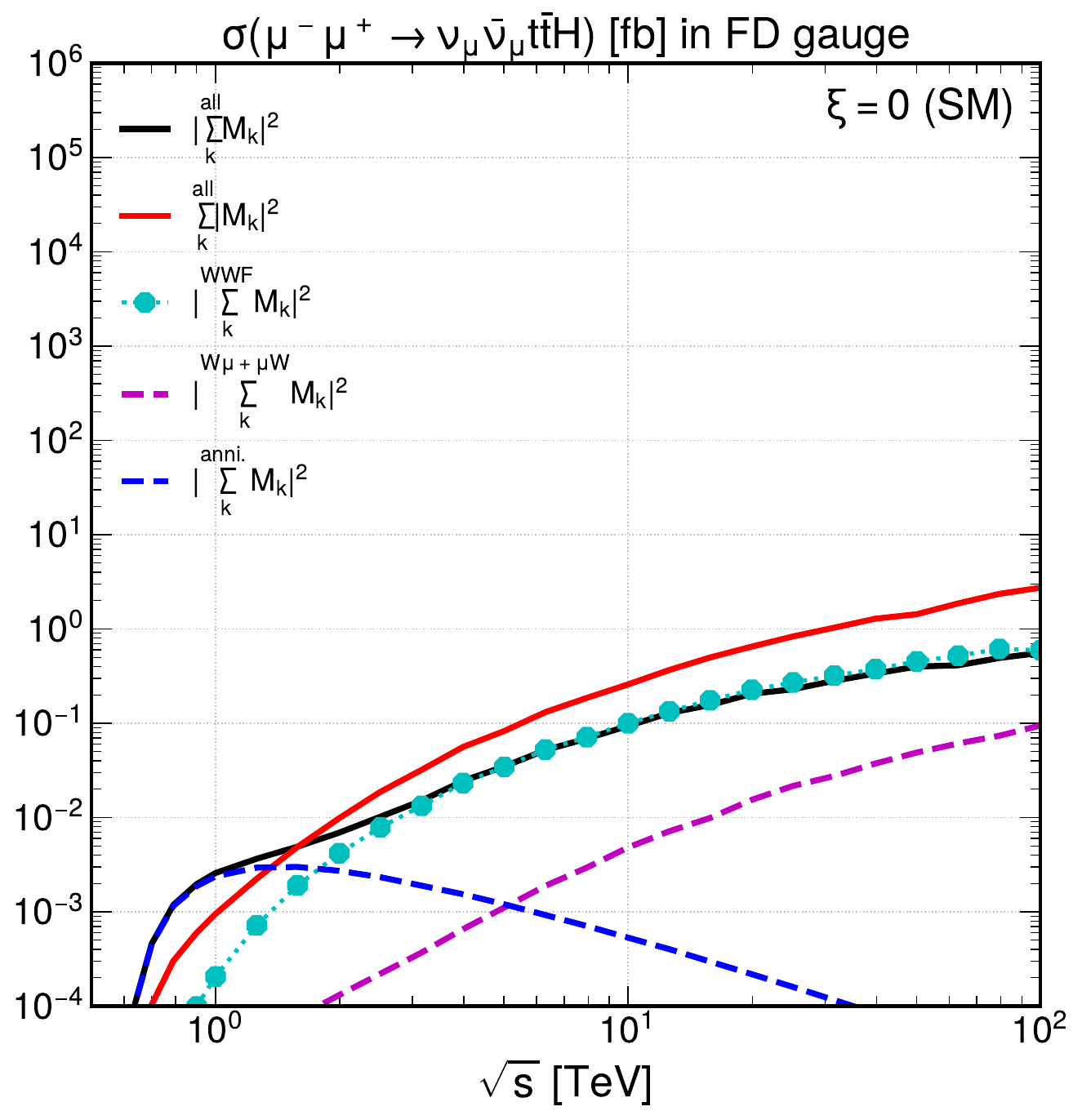}}
\caption{Total cross section of $\nu_\mu\bar{\nu}_\mu t\bar{t}H$ production with $\xi=0$ (SM) at a muon collider (black line), the sum of the squared of each amplitudes $\sum_k^{\rm all}|{\cal M}_k|^2$ (red line), the $WWF$ diagrams contribution $|\sum_{k}^{\rm WWF}{\cal M}_k|^2$ (cyan dotted),  and the sum of the three annihilation diagrams $|\sum_k^{\rm anni.}{\cal M}_k|^2$ (blue-dashed line). }
\label{fig-1}       
\end{figure}

\begin{figure}[b]
\centering
\subfigure[]{
\includegraphics[width=5cm,clip]{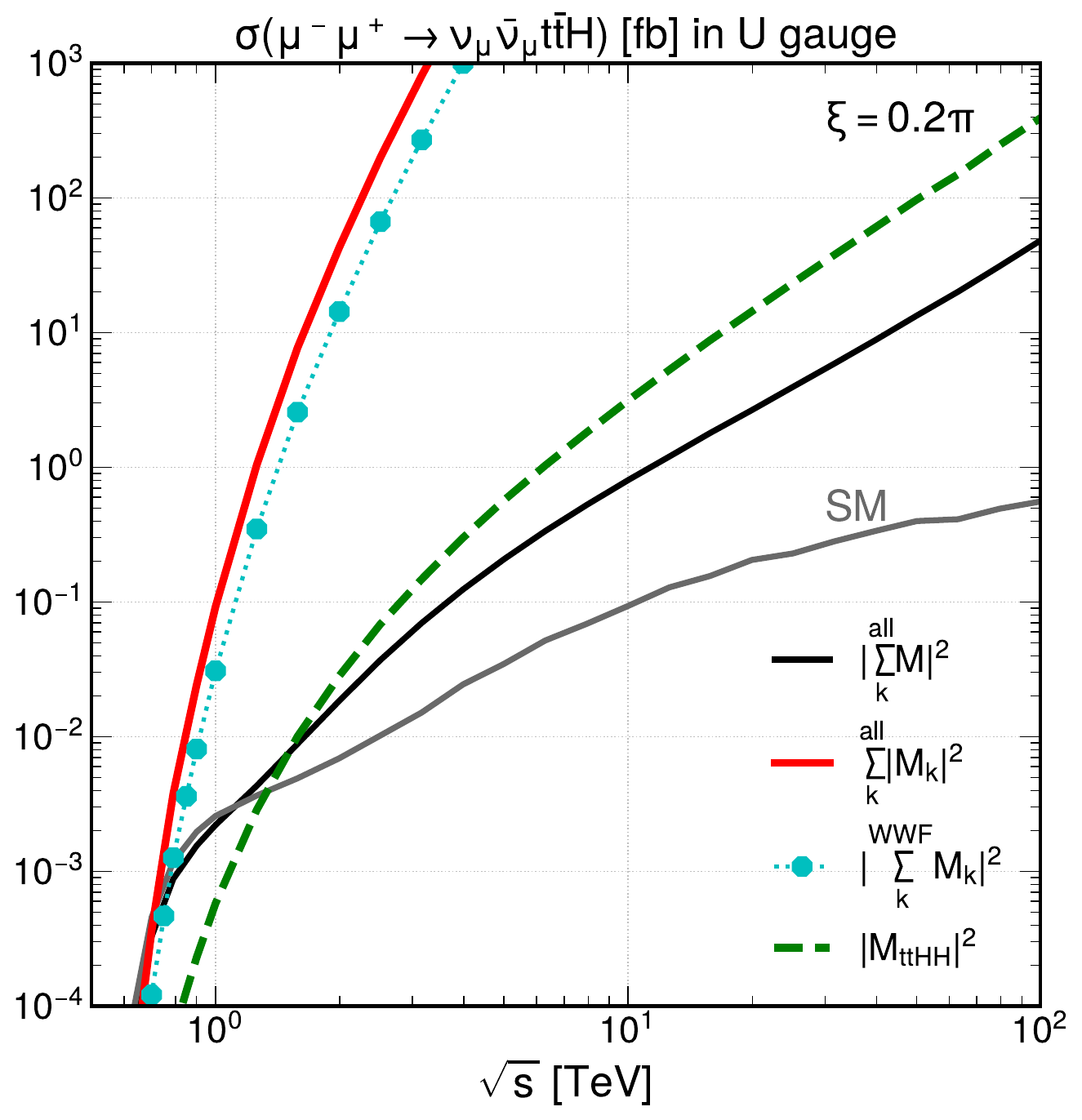}}
\subfigure[]{
\includegraphics[width=5cm,clip]{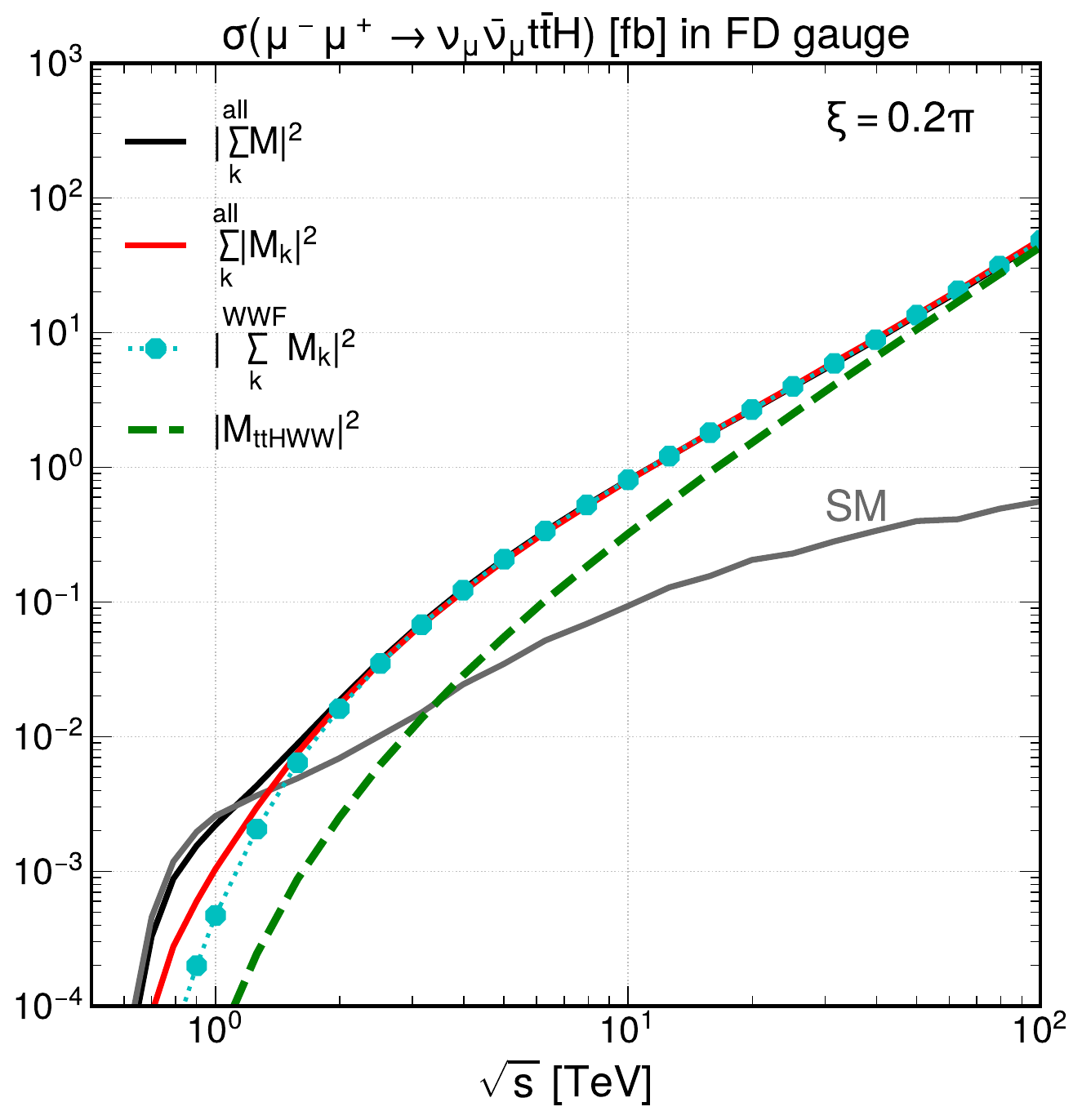}}
\caption{Total cross section of $\nu_\mu\bar{\nu}_\mu t\bar{t}H$ production with $\xi=0.2\pi$ at a muon collider (black line), the sum of the squared of each amplitudes $\sum_k^{\rm all}|{\cal M}_k|^2$ (red line), the $WWF$ diagrams contribution $|\sum_{k}^{\rm WWF}{\cal M}_k|^2$ (cyan dotted).
Single diagram contribution of $|{\cal M}_{ttHH}|^2$ in the U gauge (a) and $|{\cal M}_{ttHWW}|^2$ in the FD gauge (b) are shown by green dashed curves.
Grey line shows the SM as a reference. }
\label{fig-2}       
\end{figure}
 
 We calculate the cross sections of the process as well as the contributions of diagrams from each of the above categories. In addition, we define 
\begin{eqnarray}
    R=\frac{\sum_{helicities}\int d\Phi\sum_k|{\cal M}_k|^2}{\sum_{helicities}\int d\Phi|\sum_k{\cal M}_k|^2},
\end{eqnarray}
as a measure of subtle cancellation among interfering amplitudes, corresponding to the ratio of the red line over the black line in Fig.\ref{fig-1}.
In the unitary (U) gauge, Fig.\ref{fig-1}(a) , the contributions from the $WWF$ and $W\mu+\mu W$ type diagrams are almost degenerate in the whole energy region and their magnitudes are far larger than the total cross section.
It is after almost perfect cancellation among those two types of diagrams, the remnant gives the observable cross section. In contrast, 
for the FD gauge shown in Fig.\ref{fig-1}(b), the dominance of the $WWF$ type diagrams is clearly seen at  $\sqrt{s}\gtrsim3$ TeV.

As a non-SM example, we show in Fig.\,\ref{fig-2} the results for CP violating phase of $\xi=0.2\pi$. Fig.\,\ref{fig-2}(a) gives results in the U gauge, 
both the squared sum of the amplitudes (red line), and the $WWF$ contributions (cyan dotted) grow rapidly with energy,
just like in the SM.
In the U gauge, only one additional $ttHH$ coupling 
\begin{eqnarray}
    {\cal L}_{ttHH}
    =\frac{3(g_{\rm SM}-ge^{i\epsilon})}{2v} t_L^\dagger t_R H^2+{\rm H.c.},
\end{eqnarray}
in the SMEFT Lagrangian~\eqref{eq:LSMEFT}
coupling contributes to the process. Its contribution is shown by green-dashed line in the Fig.\,\ref{fig-2}(a). 
In the FD gauge, Fig.\,\ref{fig-2}(b), the contribution of the dimension-6 $ttHWW$ vertex is shown in green dashed line. We can tell that the total cross section is dominated by the $WWF$ contribution at $\sqrt{s}\gtrsim3$ TeV, and by $|{\cal M}_{ttHWW}|^2$ at $\sqrt{s}\gtrsim100$ TeV. This is an example that the Goldstone boson equivalence theorem is manifest in the FD gauge~\cite{Wulzer:2013mza,Chen:2016wkt}, since the $ttHWW$ vertex has only the fifth (Goldstone boson) component of the $W$'s.

We summarize our findings as follows.
In the U gauge, physical cross sections are obtained after an almost perfect cancellation among amplitudes at high energies. Consequently, it becomes challenging to identify the subamplitudes that deviate from the SM. In the FD gauge, we find no unphysical cancellation among amplitudes, and the dominance of the $WWF$ type amplitudes is clear both in the SM and also in the presence of the non-SM vertices. In addition, the FD gauge amplitudes satisfy naive scaling law and therefore the total cross section of the process $\mu^-\mu^+\to {\nu_\mu}\bar{\nu}_\mu t\bar{t}H$ at highest energies is dominated by a single diagram with the dimension-6 $ttHWW$ coupling.

\section*{Acknowledgement}
I would like to thank all the collaborators of the papers [3,6,8], V. Barger, J. Chen, K. Hagiwara, J. Kanzaki, O. Mattelaer, and K. Mawatari.
The work was supported in part by JSPS KAKENHI Grant No. 21H01077 and 23K03403.

 \bibliography{FDd6} 
%
%
%
%

\end{document}